\documentclass[reprint,preprintnumbers,nofootinbib,amsmath,amssymb,aps,prl,floatfix]{revtex4-1}
\usepackage{graphicx}
\usepackage{stmaryrd}
\usepackage{bm}
\usepackage{xcolor}

\newcommand{\blnu}{$B^-\to\ell^-\bar\nu_\ell$}
\newcommand{\nbsl}{\rlap{\hspace{0.25mm}/}{\bar n}}
\newcommand{\spac}{{\hspace{0.3mm}}}

\begin{document}
\preprint{MITP-22-111, TUM-HEP-1449/22}
% arXiv:2212.14430
% v1: December 29,2022
% v2: February 22,2023
% v3: May 26,2023

\title{Structure-Dependent QED Effects in Exclusive $\bm{B}$ Decays at Subleading Power}

\author{Claudia Cornella$^a$}
\email{claudia.cornella@uni-mainz.de}
\author{Matthias König$^b$}
\email{matthias.koenig@tum.de}
\author{Matthias Neubert$^{a,c}$}
\email{matthias.neubert@uni-mainz.de}

\affiliation{${}^a$PRISMA$^+$\! Cluster of Excellence {\em\&} MITP, Johannes Gutenberg University, 55099 Mainz, Germany\\
${}^b$Physik Department T31, Technische Universit\"at M\"unchen, James-Franck-Str.~1, 85748 Garching, Germany\\
${}^c$Department of Physics {\em\&} LEPP, Cornell University, Ithaca, NY 14853, U.S.A.}

\begin{abstract}
We derive a factorization theorem for the structure-dependent QED effects in the weak exclusive process $B^-\to\mu^-\bar\nu_\mu$, i.e., effects probing the internal structure of the $B$ meson. The derivation requires a careful treatment of endpoint-divergent convolutions common to subleading-power factorization formulas. We find that the decay amplitude is sensitive to two- and three-particle light-cone distribution amplitudes of the $B$ meson as well as to a new hadronic quantity $F(\mu,\Lambda)$, which generalizes the notion of the $B$-meson decay constant in the presence of QED effects. This is one of the first derivations of a subleading-power factorization theorem in which the soft functions are non-perturbative hadronic matrix elements.
\end{abstract}
\maketitle

Exclusive $B$-meson decays are powerful probes of the flavor sector and of physics beyond the Standard Model. In order to match the increasing experimental accuracy in several decay channels, a reliable assessment of QED corrections is desirable. In recent years, these have received considerable attention, especially in the context of leptonic and semi-leptonic $B$ decays. In most cases, QED corrections were treated via the inclusion of soft-photon emissions, under the hypothesis that the leading corrections can be described by photons unable to probe the internal meson structure \cite{Isidori:2020acz,Zwicky:2021olr}. This assumption is in direct contradiction with the observation that structure-dependent QED corrections constitute an important contribution to the decays $B_{d,s}\to\mu^+\mu^-$ \cite{Beneke:2017vpq,Beneke:2019slt}.

In this work, we present the factorization formula for the exclusive decay \blnu{} including virtual one-loop QED corrections. This process can be used to determine the CKM matrix element $V_{ub}$ and to test lepton-flavor universality, as Belle~II can perform accurate measurements of the $\ell=\mu,\tau$ channels \cite{Belle-II:2018jsg}. We focus here on the case $\ell=\mu$. Due to the chirality-suppressed nature of the decay, this process is of next-to-leading power (NLP) in the $1/m_B$ expansion. Factorization formulas at subleading power are plagued by endpoint-divergent convolution integrals \cite{Ebert:2018gsn,Moult:2019mog,Beneke:2019kgv,Moult:2019uhz,Beneke:2019oqx,Moult:2019vou,Liu:2019oav,Beneke:2020ibj,Liu:2020wbn,Liu:2020tzd,Beneke:2022obx,Bell:2022ott}, requiring a careful subtraction and rearrangement between different contributions. The ``refactorization-based subtraction (RBS) scheme'' introduced in \cite{Liu:2019oav,Liu:2020wbn} for the derivation of the factorization theorem for the Higgs-boson decay $h\to\gamma\gamma$ via $b$-quark loops provides a %systematic 
method to deal with endpoint divergences and establish factorization at NLP. The RBS scheme has also been applied successfully to Higgs production in gluon-gluon fusion \cite{Liu:2020tzd,Liu:2022ajh} and to the ``off-diagonal gluon thrust'' in $e^+ e^-$ collisions \cite{Beneke:2022obx}. Along with \cite{Feldmann:2022ixt}, the present work applies this approach for the first time in the context of exclusive rare decays of $B$ mesons, where the necessary rearrangements involve objects that are genuinely non-perturbative, giving rise to new types of hadronic matrix elements (see \cite{Hurth:2023paz} for an application to inclusive $B$ decays). The presence of such quantities is a generic feature of exclusive $B$-meson decays at NLP.

Below the electroweak scale, the effective weak Lagrangian describing the decay \blnu{} is given by 
\begin{equation}\label{Leff}
   {\cal L}_{\rm eff} 
   = - \frac{4 G_F}{\sqrt2}\spac K_{\rm EW}(\mu)\spac V_{ub}\,
    (\bar u\spac\gamma^\mu P_L b)
    (\bar\ell\spac\gamma_\mu P_L\nu_\ell) \,.
\end{equation}
When electroweak corrections are neglected $K_{\rm EW}(\mu)=1$, and all hadronic effects are encoded in the $B$-meson matrix element of the quark current, 
\begin{equation}\label{fB}
   \langle 0|\,\bar u\spac\gamma^\mu\gamma_5\spac b\,|B^-\rangle
   = i m_B f_B\spac v^\mu \,.
\end{equation}
Here $v^\mu$ denotes the 4-velocity of the $B$ meson and $f_B$ its decay constant. The situation becomes significantly more complicated when QED effects are taken into account. In this case $K_{\rm EW}(\mu)\ne 1$ \cite{Dekens:2019ept} and the operator in \eqref{Leff} has a non-trivial scale dependence, which compensates that of $K_{\rm EW}$, given by \cite{Jenkins:2017dyc}
\begin{equation}
   \frac{d K_{\rm EW}(\mu)}{d\ln\mu}
   = Q_\ell\spac Q_u\spac\frac{3\alpha}{2\pi}\spac K_{\rm EW}(\mu) \,.
\end{equation}
More profoundly, the $B$-meson decay constant loses its universal meaning and its definition must be generalized, because the flavor-changing quark current is not gauge invariant with respect to QED interactions \cite{Beneke:2019slt,Beneke:2020vnb}. The simple factorization of the four-fermion operator into a quark and a lepton current, with no interactions between them, no longer holds. While in QCD physical states are color neutral, both the $B$ meson and the charged lepton carry electric charges, and thus electromagnetic interactions inevitably connect the two currents. 

In the presence of QED effects, the $B^-\to\ell^-\bar\nu_\ell$ matrix element of the four-fermion operator in \eqref{Leff} is sensitive to six different energy scales. The first four are the scale $m_b$ setting the large mass of the decaying $B$ meson, the intermediate ``hard-collinear'' scale $\sqrt{m_b\spac\Lambda_{\rm QCD}}$ at which the internal structure of the meson is probed by virtual photons, the scale $\Lambda_{\rm QCD}$ of non-perturbative soft QCD interactions in the meson, and the lepton mass $m_\ell$. In order to obtain an infrared (IR) safe observable, it is necessary to define the decay rate for the process $B^-\to\ell^-\bar\nu_\ell\spac(\gamma)$, allowing for the emission of real photons with energies below a resolution scale $E_s$. The threshold $E_s$ and a related scale $(m_\ell/m_B)\spac E_s$ complete the list of relevant scales. We have analyzed the factorization of these scales using a multi-step procedure, in which the effective weak Lagrangian \eqref{Leff} is matched onto two versions of soft-collinear effective theory \cite{Bauer:2000yr,Bauer:2001yt,Bauer:2002nz,Beneke:2002ph}: ${\cal L}_{\rm eff}\to\text{SCET-1}\to\text{SCET-2}$. In a final step, the SCET-2 operators are matched onto a low-energy effective theory consisting of products of Wilson lines, which are needed to account for real photon emissions. Technical details will be presented elsewhere.

In this Letter, we focus on the intricate factorization properties of the decay amplitude above the scale $E_s$, which is sensitive to virtual photon exchange only. We have established the factorization theorem
\begin{equation}\label{fact}
   {\cal A}_{B\to\ell\bar\nu}^{\rm virtual}
   = \sum_j H_j\spac S_j\spac K_j 
    + \sum_i H_i\otimes J_i\otimes S_i\otimes K_i \,,
\end{equation}
where the hard functions $H_i$ account for matching corrections at the scale $m_b$, the jet functions $J_i$ encode matching corrections at the scale $\sqrt{m_b\spac\Lambda_{\rm QCD}}$, and the soft functions $S_i$ are hadronic matrix elements of the $B$ meson defined in heavy-quark effective theory (HQET) \cite{Eichten:1989zv,Georgi:1990um,Grinstein:1990mj,Neubert:1993mb}. The collinear functions $K_i$ describe the leptonic matrix elements, encoding the dependence on the scale $m_\ell$. The first set of terms arises from SCET-1 operators containing a soft spectator quark, whereas the second set descents from operators in which the spectator quark is described by a hard-collinear field, carrying a significant fraction of the charged-lepton momentum. The symbol $\otimes$ indicates that the products of component functions must be understood as convolutions, since some of the functions share common momentum variables, over which one must integrate. In SCET-2, interactions between soft 
and collinear particles can be eliminated at the Lagrangian level using field redefinitions \cite{Bauer:2001yt,Beneke:2002ni,Becher:2003qh,Fleming:2007qr}. The remnants of these interactions appear in the form of soft 
Wilson lines $Y_n^{(f)}$ for fermion $f$, which depend on its color and electric charge. The light-like vector $n$ is aligned with the direction of the muon. Soft-collinear photons, whose momenta are collinear with the muon but whose energy in the $B$-meson rest frame is smaller than the soft scales $\Lambda_{\rm QCD}$ and $E_s$ by a factor $m_\ell/m_B$, also play an important role. Removing their interactions with (massive) collinear and soft particles by field redefinitions gives rise to soft-collinear Wilson lines $C_{v_\ell}^{(\ell)}$ for the muon and $C_{\bar n}^{(u)\dagger} C_{\bar n}^{(b)}\equiv C_{\bar n}^{(B)}$ for the two valence quarks inside the $B$-meson. Here $v_\ell$ denotes the 4-velocity of the lepton, and the light-like vector $\bar n$ is aligned with the direction of the neutrino. These soft-collinear Wilson lines are inherited by the effective theory below the scale $\Lambda_{\rm QCD}$. For the purposes of our discussion here, they can simply be ignored.

\begin{figure}
\centering
\includegraphics[scale=0.7]{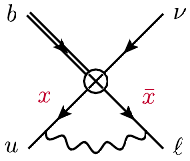}\qquad\qquad
\includegraphics[scale=0.7]{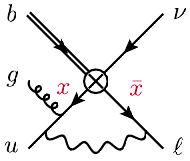}
\caption{SCET-1 loop diagrams generating structure-dependent QED corrections. The up-quark and muon leaving the weak-interaction operator carry fractions $x$ and $\bar x=1-x$ of the large component $\bar n\cdot p_\ell$ of the muon momentum. The resulting contributions involve convolutions with a two-particle (left) and three-particle (right) LCDA of the $B$ meson.}
\label{fig:hcquark}
\vspace{-1mm}
\end{figure}

The appearance of a hard-collinear scale between $m_b$ and $\Lambda_{\rm QCD}$ is an important feature of the factorization formula. Electromagnetic radiation with virtuality $q^2\sim m_b\spac\Lambda_\mathrm{QCD}$ emitted from the muon can recoil against the meson and probe its internal structure. This effect arises from the interactions between soft and collinear particles \cite{Beneke:1999br,Beneke:2000ry,Beneke:2001ev}, which in SCET-1 are mediated by the exchange of a virtual photon between the muon and the soft spectator quark in the $B$ meson, as illustrated in Figure~\ref{fig:hcquark}. After matching onto SCET-2 this gives rise to non-local operators, whose component fields have light-like separation. Their matrix elements define the $B$-meson light-cone distribution amplitudes (LCDAs) \cite{Grozin:1996pq,Lange:2003ff,Kawamura:2001jm,Braun:2017liq}. From a systematic analysis of the operators contributing to the decay at $\mathcal{O}(1/m_b)$, we find that the amplitude is sensitive to a hadronic parameter $F$ generalizing the concept of the $B$-meson decay constant, as well as to two- and three-particle LCDAs of the $B$ meson.

A natural definition of the parameter $F$ would be in terms of the $B$-meson matrix element of the operator  
\begin{equation}\label{OAdef}
   O_A = \bar n_\mu\spac\bar u_s \gamma^\mu P_L h_v\spac Y_n^{(\ell)\dagger} \,,
\end{equation}
where $u_s$ denotes a soft quark field, and $h_v$ is the effective $b$-quark field in HQET. The factor $\bar n_\mu$ appears in the evaluation of the leptonic matrix element. The Wilson line arises from the decoupling of soft photons from the muon. It ensures that the operator is gauge invariant under both QCD and QED. In the presence of QED corrections, the anomalous dimension of $O_A$ exhibits a sensitivity to IR regulators, which must be removed with a suitable subtraction \cite{Beneke:2019slt,Beneke:2020vnb}. Following these authors, one can thus define $F$ as the matching coefficient of the $B$-meson matrix element of $O_A$ onto a Wilson-line operator in a low-energy effective theory for very soft photons (with $E_\gamma \ll \Lambda_{\rm QCD}$), which see the $B$ meson as a point-like particle:
\begin{equation}\label{notyet} 
   \langle 0|\spac O_A |B^-\rangle
   = - \frac{i}{2}\spac\sqrt{m_B}\,F(\mu) \,
    \langle 0|\spac Y_{v}^{(B)} Y_n^{(\ell)\dagger}\spac|0\rangle \,. 
\end{equation}
However, an unusual aspect of this definition is that the renormalization of the ``local" (with regard to the quark fields) operator $O_A$ requires the non-local operator
\begin{equation}\label{OBdef}
  O_B(\omega) 
   = \int\frac{dt}{2\pi}\, e^{i\omega t}\,
    \bar u_s(tn)\spac[tn,0]\,\nbsl\spac P_L h_v(0)\,Y_n^{(\ell)\dagger}(0)  
\end{equation}
as a counterterm. Here the quark fields are separated by a light-like distance, and $[tn,0]$ denotes a soft Wilson-line segment connecting them.

There exists another problem with the factorization formula \eqref{fact}, as some of the convolution integrals suffer from endpoint divergences. This is a common feature of NLP factorization theorems. Neglecting corrections of $\mathcal{O}(\alpha\spac\alpha_s)$, the divergent convolutions are those involving the hard and jet functions. These divergences are troublesome, because they give rise to $1/\epsilon$ poles that cannot be removed by renormalizing the hard and jet functions individually, and hence break the desired factorization of scales. Interestingly, we find that removing the endpoint divergences using the RBS scheme \cite{Liu:2019oav,Liu:2020wbn} allows us to redefine the soft operator $O_A$ in such a way that it no longer mixes with the non-local operator $O_B(\omega)$.

The RBS scheme offers a systematic procedure for dealing with endpoint divergences. In a first step, they are removed by performing plus-type subtractions of the integrand, i.e.\
\begin{equation}
\begin{aligned}
   & H_i\otimes J_i \equiv \int_0^1\!dx\,H_i(m_b,x)\,J_i(m_b\spac\omega,x) \\
   &\to \int_0^1\!dx\spac\Big[ H_i(m_b,x)\,J_i(m_b\spac\omega,x) \\
   &\hspace{1.55cm} - \theta(\lambda-x)\,
    \llbracket H_i(m_b,x)\rrbracket\spac
    \llbracket J_i(m_b\spac\omega,x)\rrbracket \Big] \,,
\end{aligned}
\end{equation}
where $x$ is a shared longitudinal momentum fraction defined in Figure~\ref{fig:hcquark}, and $\omega$ denotes the $n\cdot p_u$ component of the soft spectator momentum, which the jet and soft functions share. (In some cases there can be more than one such variable.) The singular limit is $x\to 0$, corresponding to the region in which the virtual spectator quark becomes soft. The double brackets indicate that one needs to retain only the leading singular terms in the expressions for the hard and jet functions. More accurately, when $x=\mathcal{O}(\Lambda_{\rm QCD}/m_b)$ the quark and photon propagators in the loop become soft and should no longer be described using hard-collinear fields. We introduce a parameter $0<\lambda<1$ to subtract these contributions (see also \cite{Beneke:2022obx}). By construction, this subtraction removes the endpoint divergence, but the subtraction term must be added back in a consistent way. This is done using exact, $D$-dimensional refactorization conditions \cite{Liu:2019oav,Liu:2020wbn,Beneke:2020ibj}, which govern the structure of the component functions in the singular limits. In our case, these conditions read
\begin{equation}
\begin{aligned}
  \llbracket H_i(m_b,x)\rrbracket 
   &= H_i'(m_b)\,S_i'(\omega') \,, \\[1mm]
   \llbracket J_i(m_b\spac\omega,x)\rrbracket
   &= m_b\,S_i''(\omega,\omega') \,,
\end{aligned}
\end{equation}
where $H_i'$ are new hard functions, while $S_i'$, $S_i''$ are new soft functions, which depend on the variable $\omega'\equiv x\spac m_b$. The term that needs to be added back thus takes the form of a hard matching coefficient times a soft function, 
\begin{equation}\label{rearrange}
\begin{aligned}
   & \int d\omega \int_0^\lambda\!dx\,\llbracket H_i(m_b,x)\rrbracket\spac 
    \llbracket J_i(m_b\spac\omega,x)\rrbracket\spac S_i(\omega) \\
   &= - H_i' \int d\omega \int_\Lambda^\infty\!d\omega'\spac 
    \hat S_i(\omega,\omega') \,,
\end{aligned}
\end{equation}
where $\Lambda=\lambda\spac m_b$. In the last step we have defined $\hat S_i=S_i\spac S_i'\spac S_i''$ and added a scaleless integral, which vanishes in dimensional regularization. After adding back this term, it can be combined with other terms of similar form. 

Let us illustrate this procedure for the soft operators relevant for the subtraction of endpoint divergences in our problem. These are the local operator $O_A$ in \eqref{OAdef} and the associated non-local operator $O_B(\omega)$ in \eqref{OBdef}. There is also a third operator giving rise to a three-particle LCDA, which we omit here for simplicity, but we include its effect in our final result \eqref{master} below. (The SCET-1 and SCET-2 operator bases needed to establish the factorization theorem are of course much larger. They can be found by building all gauge- and boost-invariant operators of mass dimension~6 and the correct power counting \cite{Beneke:2002ni,Becher:2003qh}. The additional operators do not give rise to endpoint divergences, however.) The contribution of these two operators to the decay amplitude can be written as
\begin{equation}\label{ABterms}
\begin{aligned}
   & {\cal A}_{B\to\ell\bar\nu}^{(A,B)}
    = - \frac{4 G_F}{\sqrt2}\spac K_{\rm EW}\spac V_{ub}\,\frac{m_\ell}{m_b}\,
    K_A(m_\ell)\,\bar u(p_\ell) P_L\spac v(p_\nu) \\ 
   &\!\cdot\! \bigg[  H_A(m_b)\spac S_A \!+\!\! \int\!d\omega\! 
    \int_0^1\!\!dx\spac H_B(m_b,x)\spac J_B(m_b\spac\omega,x)\spac 
    S_B(\omega) \bigg] , 
\end{aligned}
\end{equation}
where $S_A=-\frac{i}{2}\sqrt{m_B}\spac F$, $S_B(\omega)=-\frac{i}{2}\sqrt{m_B}\spac F\spac\phi_-^B(\omega)$, and $H_{A,B}=1+\mathcal{O}(\alpha_s,\alpha)$. Here $\phi_-^B(\omega)$ is one of the twist-3 LCDAs of the $B$ meson \cite{Grozin:1996pq}, which is normalized to unity. Starting at one-loop order $H_B$ contains logarithmic singularities at $x=0$ ($\sim x^{-n\epsilon}$ in dimensional regularization). The collinear functions for the two contributions are equal and normalized so that $K_A=K_B=1+\mathcal{O}(\alpha)$. At one-loop order, the (bare) jet function is given by
\begin{equation}
\begin{aligned}
   J_B(m_b\spac\omega,x) 
   &= - Q_\ell\spac Q_u\spac\frac{\alpha}{2\pi}\spac
    \frac{e^{\epsilon\gamma_E}\spac\Gamma(\epsilon)}{1-\epsilon} 
    \left( \frac{1}{x} + 1 - 2\epsilon \right) \\
   &\quad\cdot \left( \frac{\mu^2}{m_b\spac\omega\spac x(1-x)}\right)^\epsilon .
\end{aligned}
\end{equation}
The refactorization conditions for $H_B$ and $J_B$ read
\begin{equation}
\begin{aligned}
   \llbracket H_B(m_b,x)\rrbracket
   &= H_A(m_b)\spac S_B'(\omega') \,, \\[1mm]
   \llbracket J_B(m_b\spac\omega,x)\rrbracket
   &= m_b\spac S_B''(\omega,\omega') \,,
\end{aligned}
\end{equation}
where $\omega'=x\spac m_b$, $S_B^\prime = 1 + \mathcal{O}(\alpha_s,\alpha)$, and
\begin{equation}
   S_B''(\omega,\omega') 
   = - Q_\ell\spac Q_u\spac\frac{\alpha}{2\pi}\spac
    \frac{e^{\epsilon\gamma_E}\spac\Gamma(\epsilon)}{1-\epsilon}\,
    \frac{1}{\omega'} \left( \frac{\mu^2}{\omega\spac\omega'}\right)^\epsilon .
\end{equation}
We find that at one-loop order the subtraction term in \eqref{rearrange} is given by 
\begin{equation}\label{subtraction}
   H_A\spac S_A\spac Q_\ell\spac Q_u\spac\frac{\alpha}{2\pi}\spac
    \frac{e^{\epsilon\gamma_E}\spac\Gamma(\epsilon)}{1-\epsilon} 
    \int_0^{\infty}\!d\omega\spac\phi_-^B(\omega)\!\int_\Lambda^\infty\frac{d\omega'}{\omega'} 
    \left( \frac{\mu^2}{\omega\spac\omega'}\right)^\epsilon\! .
\end{equation}
We consistently neglect terms of $\mathcal{O}(\alpha^2)$ and thus do not include QED corrections to the LCDA. The presence of the hard function $H_A$ in this result suggests that we should combine it with the contribution of the operator $O_A$. In all previous applications of the RBS scheme, the soft functions were perturbatively calculable, and the effect of the subtraction terms could be worked out order by order in perturbation theory. In the present case, we apply the refactorization conditions for the first time in a non-perturbative context, where the soft functions are hadronic matrix elements, which cannot be calculated using short-distance methods. Adding the subtraction term (including the three-particle contribution neglected above) has the effect of replacing the operator $O_A$ by  
\begin{equation}\label{OAlambda}
   O_A\to O_A^{(\Lambda)}    
   =\bar u_s\spac\nbsl\spac\left( 1-\theta_T(-i\bar n\cdot \overleftarrow{D}_s - \Lambda)\right) \spac  P_L h_v\,Y_n^{(\ell)\dagger}  \,, 
\end{equation}
where the covariant derivative in the $\theta$-function acts on the light quark, and the symbol ``$T$’’ on the Heaviside distribution means that in evaluating the matrix elements of this operator one needs to perform a Taylor expansion, treating $\Lambda\gg\Lambda_{\rm QCD}$ as parametrically larger than all soft QCD scales. Due to the $T$ symbol, it is not legitimate to replace the terms inside the square brackets by a Heaviside function with the opposite-sign argument. Generalizing \eqref{notyet}, we now define the hadronic parameter $F$ as
\begin{equation}\label{Fnewdef}
   \langle 0|\spac O_A^{(\Lambda)} |B^-\rangle
   = - \frac{i}{2}\spac\sqrt{m_B}\,F(\mu,\Lambda)\,
    \langle 0 |\spac Y_{v}^{(B)} Y_n^{(\ell)\dagger}\spac|0\rangle \,.
\end{equation}
We find that the presence of the $\theta$-function in \eqref{OAlambda} removes the mixing with the non-local operator $O_B(\omega)$. At one loop order, the anomalous dimension defined via $dF(\mu,\Lambda)/d\ln\mu=-\gamma_F\spac F(\mu,\Lambda)$ is given by
\begin{equation}
   \gamma_F = - C_F\spac\frac{3\alpha_s}{4\pi} 
    + \frac{3\alpha}{4\pi} \left( Q_\ell^2 - Q_b^2 + \frac23\spac Q_\ell\spac Q_u \ln\frac{\Lambda^2}{\mu^2} \right) . 
\end{equation}
It is also possible to control the dependence on the cutoff $\Lambda$ using perturbation theory. At one-loop order, we find
\begin{equation}\label{Lamdep}
   \frac{d\ln F}{d\ln\Lambda}
   = Q_\ell\spac Q_u\spac\frac{\alpha}{2\pi} \left[ \int_0^\infty\!d\omega\spac\phi_-^B(\omega) 
    \ln\frac{\omega\spac\Lambda}{\mu^2} - 1 + \dots \right] ,
\end{equation}
where the dots stand for a contribution involving the three-particle LCDA.

When the subtraction term is combined with the original contribution of the operator $O_A$, we obtain from \eqref{ABterms} 
\begin{align}\label{rearranged}
   &{\cal A}_{B\to\ell\bar\nu}^{(A,B)}
    = - \frac{4 G_F}{\sqrt2}\spac K_{\rm EW}\spac V_{ub}\,\frac{m_\ell}{m_b}\,
     S_A^{(\Lambda)} K_A(m_\ell)\,\bar u(p_\ell) P_L\spac v(p_\nu) \notag\\ 
   &\cdot \bigg[ H_A(m_b) + \int_0^{\infty}\!d\omega\spac\phi_-^B(\omega)\!
    \int_0^1\!dx\spac \Big[ H_B(m_b,x)\spac J_B(m_b\spac\omega,x) \notag\\
   &\hspace{1.1cm} - \theta(\lambda-x)\,\llbracket H_B(m_b,x)\rrbracket\spac\llbracket J_B(m_b\spac\omega,x)\rrbracket
    \Big] \bigg] \,. 
\end{align}
The subtracted convolution and the soft function $S_A^{(\Lambda)}=-\frac{i}{2}\sqrt{m_B}\spac F(\mu,\Lambda)$ depend on the cutoff $\Lambda$, and there is no choice for which both objects depend only on their natural scales. Following \cite{Liu:2019oav}, we choose $\Lambda=m_b$ and hence $\lambda=1$ to eliminate the second scale from the subtracted convolution, at the expense of introducing the scale $m_b$ in the definition of $F$ in \eqref{Fnewdef}. The translation of $F(\mu,m_b)$ to $F(\mu,\Lambda)$ with a different choice of $\Lambda$ can be obtained by solving the evolution equation \eqref{Lamdep}.

We are now ready to present our main result. We find that the $B^-\to\mu^-\bar\nu_\mu$ decay amplitude including virtual QED corrections is given by
\begin{equation}
\begin{aligned}
   \mathcal{A}_{B^\to\ell\bar\nu}^{\rm virtual} 
   &= i\sqrt2\spac G_F\spac K_{\rm EW}(\mu)\spac V_{ub}\,\frac{m_\ell}{m_b}\,\bar u(p_\ell) P_L\spac v(p_\nu) \\
   &\quad\cdot\sqrt{m_B}\spac F(\mu,m_b)\spac\Big[ \mathcal{M}_{2p}(\mu) +  \mathcal{M}_{3p}(\mu) \Big] \,,
\end{aligned}
\end{equation}
where the two terms in the second line probe the two- and three-particle Fock states of the $B$ meson. After renormalizing the four-fermion operator in \eqref{Leff}, the muon mass and the parameter $F$ in the $\overline{\rm MS}$ scheme, and performing the integrations over $x$, we obtain at one-loop order
\begin{widetext}
\begin{align}\label{master}
   \mathcal{M}_{2p}(\mu) 
   &= 1 + \frac{C_F\spac\alpha_s}{4\pi} \left[ \frac{3}{2} \ln\frac{m_b^2}{\mu^2} - 2 \right] 
    + \frac{\alpha}{4\pi} \left\{ Q_b^2 \left[ \frac{3}{2} \ln\frac{m_b^2}{\mu^2} - 2 \right] 
    - Q_\ell\spac Q_b \left[ \frac12 \ln^2\frac{m_b^2}{\mu^2} + 2\ln\frac{m_b^2}{\mu^2} - 3\ln\frac{m_\ell^2}{\mu^2} 
    + 6 + \frac{5\pi^2}{12} \right] \right. \notag\\
   &\quad \left. + 2\spac Q_\ell\spac Q_u \int_0^\infty\!d\omega\,\phi_-^B(\omega) \ln\frac{m_b\spac\omega}{\mu^2}    
    + Q_\ell^2 \left[ \frac{1}{\epsilon_{\rm IR}} \left( \ln\frac{m_B^2}{m_\ell^2} - 2 \right) 
    + \frac12 \ln^2\frac{m_\ell^2}{\mu^2} - \frac12 \ln\frac{m_\ell^2}{\mu^2} + 5 + \frac{5\pi^2}{12} \right] 
    \right\} , \\
   \mathcal{M}_{3p}(\mu)
   &= - \frac{\alpha}{\pi}\,Q_\ell\spac Q_u
    \int_0^\infty\!d\omega \int_0^\infty\!d\omega_g\,\phi_{3g}^B(\omega,\omega_g) 
    \left[ \frac{1}{\omega_g} \ln\bigg( 1 + \frac{\omega_g}{\omega} \bigg) 
    - \frac{1}{\omega+\omega_g} \right] . \notag
\end{align}
\end{widetext}
In the virtual amplitude there remain IR divergences, which cancel against the IR divergences from real-photon emission in the process $B^-\to\ell^-\bar\nu_\ell\spac(\gamma)$.

When corrections of $\mathcal{O}(\alpha\spac\alpha_s)$ are included, the integrals over the LCDA $\phi_-^B(\omega)$ in \eqref{rearranged} and \eqref{master} no longer converge at infinity \cite{Grozin:1996pq,Lange:2003ff}, indicating another occurrence of an endpoint divergence. One then needs to refactorize these integrals in the region where $\Lambda_{\rm QCD}\ll\omega\ll m_b$, using techniques developed in \cite{Lee:2005gza,Beneke:2023nmj}. Since these corrections are bound to be very small numerically, this issue will be discussed in detail elsewhere.

The three-particle LCDAs of the $B$ meson have been studied in \cite{Kawamura:2001jm,Braun:2017liq}. Our function $\phi_{3g}^B(\omega,\omega_g)$ is related to the functions defined in these references by
\begin{equation}
   \phi_{3g}^B(\omega,\omega_g)
   = \frac{1}{\omega_g}\spac \Big[ \psi_A(\omega,\omega_g) - \psi_V(\omega,\omega_g) \Big] \,,
\end{equation}
where the momentum variables $\omega$ and $\omega_g$ refer to the spectator quark and the gluon, respectively. For small values of these parameters one finds the asymptotic behavior $\phi_{3g}^B(\omega,\omega_g)\propto\omega\spac\omega_g$ \cite{Braun:2017liq}, showing that the convolution integral in the three-particle term is convergent.

The above expressions show the structure-dependent nature of the QED corrections. The appearance of the two- and three-particle LCDAs highlights the fact that hard-collinear photons are energetic enough to probe the internal structure of the $B$ meson. Various phenomenological models for the LCDAs have been proposed in the literature \cite{Grozin:1996pq,Kawamura:2001jm,Braun:2017liq} and could be used to obtain an estimate of these effects. The terms sensitive to the quark electric charges in \eqref{master} are missed in a theory in which the $B$ meson is treated as a point-like particle. It is evident that the one-loop QCD corrections contain large single and double logarithms, which can be resummed using renormalization-group equations in SCET. 

In the absence of QED corrections, we have
\begin{equation}
   \sqrt{m_B}\spac f_B^{\rm QCD}
   = \left[ 1 - C_F\spac\frac{\alpha_s(m_b)}{2\pi} \right] F(m_b,m_b)
    \Big|_{\alpha\to 0} 
\end{equation}
up to power corrections of $\mathcal{O}(1/m_b)$. The parameter $f_B^{\rm QCD}$ can be computed with high precision using lattice QCD \cite{HFLAV:2022pwe}. While the QED correction included in the definition of $F$ is expected to be small, being governed by $\alpha$, its value is sensitive to non-perturbative dynamics and difficult to estimate. Due to the presence of the light-like Wilson line in \eqref{Fnewdef}, it appears challenging to compute $F$ on a Euclidean lattice.

To summarize, we have derived the first QCD\,+\,QED factorization formula for a NLP observable using SCET methods. Focusing on the virtual QED corrections to the exclusive $B^-\to\mu^-\bar\nu_\mu$ decay amplitude, our main goal was to separate perturbative QED corrections from non-perturbative ones, which are sensitive to hadronic dynamics. This is of great importance for future precision determinations of the CKM matrix element $V_{ub}$, because the new hadronic parameter $F(\mu,\Lambda)$ and the $B$-meson LCDAs introduce significant hadronic uncertainties in the analysis of QED corrections. Our derivations have required a careful handling endpoint-divergent convolutions, which we have treated in the RBS scheme \cite{Liu:2019oav,Liu:2020wbn}. While this scheme has previously been applied to other observables, our case is special in that applying refactorization in a non-perturbative context requires a modification of the relevant hadronic matrix elements. This leads to the introduction of the $\theta$-function in \eqref{OAlambda} and thus to a novel class of soft operators. The result \eqref{master} is valid for $\ell=\mu$ only, and its generalization to other lepton flavors will be discussed elsewhere. The approach presented here provides a framework for future studies of structure-dependent QED corrections to other rare exclusive $B$ decays at NLP.
\\[-2mm]

\paragraph*{Acknowledgments.}
We are grateful to Martin Beneke, Philipp B\"oer, and Jian Wang for useful comments, and to Gino Isidori for suggesting to study this problem. C.C.\ and M.K.\ thank the Mainz Institute for Theoretical Physics for hospitality during the program {\em Power Expansions on the Light-Cone} (September 19--30, 2022). The research of C.C.\ and M.N.\ was supported by the Cluster of Excellence \textit{Precision Physics, Fundamental Interactions, and Structure of Matter} (PRISMA$^+$, EXC 2118/1) within the German Excellence Strategy (Project-ID 39083149). M.K.\ was supported by the German Research Foundation (DFG) through the Sino-German Collaborative Research Center TRR~110 \textit{Symmetries and the Emergence of Structure in QCD} (Project-ID 196253076, NSFC Grant No.~12070131001). C.C.\ is grateful for the hospitality of Perimeter Institute (PI), where part of this work was carried out. Research at PI is supported in part by the Government of Canada through the Department of Innovation, Science and Economic Development Canada and by the Province of Ontario through the Ministry of Economic Development, Job Creation and Trade. C.C.\ was also supported in part by the Simons Foundation through the Simons Foundation Emmy Noether Fellows Program at PI. 

\vspace{-2mm}
\bibliography{bibliography}

\end{document}